\documentclass[twocolumn]{article}

\usepackage{xcolor}
\usepackage{amsmath,amsfonts,amssymb}
\usepackage{algorithmic}
\usepackage{algorithm}
\usepackage{array}
\usepackage[caption=false,font=normalsize,labelfont=sf,textfont=sf]{subfig}
\usepackage{textcomp}
\usepackage{stfloats}
\usepackage{url}
\usepackage{verbatim}
\usepackage{graphicx}
\usepackage{cite}
\usepackage{soul}
\usepackage{mathtools}
\usepackage{geometry}
\usepackage{titlesec}
\usepackage{fancyhdr}

\titlespacing*{\section}{0pt}{12pt}{6pt}
\titlespacing*{\subsection}{0pt}{10pt}{5pt}
\titlespacing*{\subsubsection}{0pt}{8pt}{4pt}

\newcommand{\real}{\ensuremath{\mathbb{R}}}

\newcommand{\R}{\mathbb{R}}

\newcommand{\Sym}{\mathsf{Sym}}

\newtheorem{theorem}{Theorem}
\newtheorem{corollary}{Corollary}
\newtheorem{definition}{Definition}
\newtheorem{proposition}{Proposition}

\newenvironment{proof}{\textit{Proof.}}{\hfill$\square$}

\definecolor{darkred}{rgb}{0.8, 0.0, 0.0}

\begin{document}

\title{Fused Gromov-Wasserstein Variance Decomposition with Linear Optimal Transport}
\author{Michael Wilson, Tom Needham, and Anuj Srivastava}
\date{}
\maketitle

\begin{abstract}
Wasserstein distances form a family of metrics on spaces of probability measures that have recently seen many applications.  
However, statistical analysis in these spaces is complex due to the nonlinearity of Wasserstein spaces. One potential solution to this problem is Linear Optimal Transport (LOT). This method allows one to find a Euclidean embedding, called {\it LOT embedding}, of measures in some Wasserstein spaces, but some information is lost in this embedding. So, to understand whether statistical analysis relying on LOT embeddings can make valid inferences about original data, it is helpful to quantify how well these embeddings describe that data. To answer this question, we present a decomposition of the {\it Fr\'echet variance} of a set of measures in the 2-Wasserstein space, which allows one to compute the percentage of variance explained by LOT embeddings of those measures. We then extend this decomposition to the Fused Gromov-Wasserstein setting. We also present several experiments that explore the relationship between the dimension of the LOT embedding, the percentage of variance explained by the embedding, and the classification accuracy of machine learning classifiers built on the embedded data. We use the MNIST handwritten digits dataset, IMDB-50000 dataset, and Diffusion Tensor MRI images for these experiments. Our results illustrate the effectiveness of low dimensional LOT embeddings in terms of the percentage of variance explained and the classification accuracy of models built on the embedded data.
\end{abstract}

% \tableofcontents{}

\section{Introduction}

Probability measures are objects of fundamental importance in statistics, and Optimal Transport (OT) theory~\cite{villani2009optimal} provides a powerful framework for studying them. Recently, computational methods for optimal transport~\cite{COTFNT, cuturi2013sinkhorn} have found diverse applications in machine learning and data science~\cite{montesuma2023recent, chewi2024statistical}. These include applications in computer vision~\cite{boneel2023survey}, as well as in generative modeling~\cite{arjovsky2017wasserstein, genevay2018learning, sander2022sinkformers} and domain adaptation~\cite{courty2016optimal}. Variants of OT have also been applied to diffeomorphic image registration with Wasserstein-Fisher Rao metrics~\cite{chizat2018interpolating, feydy2017optimal, feydy2019fast}, comparison of Gaussian mixture measures~\cite{delon2020wasserstein,wilson2024wasserstein}, and shape or graph matching with Gromov-Wasserstein and Fused Gromov-Wasserstein distances~\cite{memoli2007use, memoli2011gromov, chowdhury2019gromov, vayer2018fused,vayer2019optimal,chowdhury2020gromov,chowdhury2021generalized}.

This paper focuses on the statistical analysis of datasets where each observation is a point cloud or graph, represented as an empirical probability measure in a metric space. A significant challenge in this approach is that such data objects have no canonical Euclidean representation and thus do not lend themselves to statistical techniques designed for Euclidean data. One potential solution to this problem, called {\it Linear Optimal Transport} (LOT) \cite{Wang2013}, offers an embedding of measures supported on $\R^d$ (and even some Riemannian manifolds~\cite{sarrazin2023linearized}) into a given Euclidean space. However, given that the original data are not Euclidean, one should expect some information to be lost in the embedding process. Thus, to determine whether LOT should be used on a particular dataset, it would be helpful to quantify how much information is lost, how this relates to the dimension of the embedding, and what consequences this has on subsequent statistical analysis. 

To address these questions, we present a decomposition of {\it Fr\'echet variance}~\cite{dubey2019frechet} in the 2-Wasserstein space, which allows one to compute the percentage of variance captured by Linear Optimal Transport embeddings. This is similar in spirit to the non-Euclidean PCA (e.g., tangent space PCA~\cite{fletcher2004principal}) where one evaluates a PCA representation using the fraction of total variance captured by the linear approximation. We then proceed to show that similar decompositions of Fr\'echet variance can be found for the Gromov-Wasserstein (GW) and Fused Gromov-Wasserstein (FGW) distances. 

This paper first recalls the Wasserstein distance, Wasserstein barycenters, and the concept of Fr\'echet variance, focusing on empirical probability measures. It then develops a principled way to decompose Fr\'echet variance in Wasserstein space and demonstrates this idea practically using several experiments on real data. We also recall the concept of (Fused) Gromov-Wasserstein distance and show how our results on Fr\'echet variance and our numerical framework extend to this setting. Our  experiments study the relationships between the dimension of the LOT embedding, the percentage of variance explained by the embedding, and the accuracy of machine learning classifiers built on the embedded data. We apply these experiments to the MNIST handwritten digits dataset~\cite{mnist}, IMDB-50000 sentiment analysis dataset \cite{maas2011learning}, and Diffusion Tensor MRI (DTMRI) data from the ``Human Connectome Project - Young Adult'' (HCP-YA) dataset~\cite{van2012human}. Our results show that LOT can be used to find low-dimensional Euclidean embeddings of unstructured data while maintaining relatively high classification accuracy, especially in applications to computer vision. 

The specific contributions of this work are as follows.
\begin{enumerate}
    \item Using a novel interpretation of a decomposition of squared 2-Wasserstein distances previously noted in \cite{alfonsi2020squared, bai2023linear}, we express the 2-Wasserstein Fr\'echet variance as a sum of deterministic and probabilistic components.  
    
    \item We introduce a similar decomposition of the 2-Gromov-Wasserstein Fr\'echet variance into deterministic and probabilistic components, including a proof that the deterministic component in the decomposition corresponds to a 2-Gromov-Wasserstein Fr\'echet variance. Using these results, we show that one can also define a decomposition of Fused 2-Gromov-Wasserstein Fr\'echet variance. 

    \item We present a novel extension of the $F$-statistic that allows one to test for equality of $n$ support barycentric projection for point clouds, with accompanying examples that show how to conduct permutation tests using the statistic on simulated data. 
    
    \item We present experiments that show how LOT and the variance decomposition can be used for dimensionality reduction of point clouds and graphs and for parameter selection in the Fused Gromov-Wasserstein setting. Our results illustrate that low dimensional LOT embeddings can capture much of the Fr\'echet variation and provide relatively good classification results in real-world datasets.
\end{enumerate}

The rest of this paper is organized as follows; in Section \ref{sec: background}, we present relevant background on Optimal Transport and Linear Optimal Transport. In Section \ref{sec: novel results}, we present a decomposition of the 2-Wasserstein Fr\'echet variance for datasets of empirical probability measures supported on $\R^d$ into deterministic and probabilistic components, and then extend this decomposition to the Gromov-Wasserstein and Fused Gromov-Wasserstein settings. In Section \ref{sec: numerical experiments}, we present the results of numerical experiments that explore the variance decompositions using the MNIST handwritten digits dataset, natural language data from the IMDB-50000 dataset, and DTMRI data from the HCP-YA dataset. In Section \ref{Conclusion}, we conclude by discussing directions for future work.

\section{Background Material} \label{sec: background}

\subsection{Wasserstein Distances and Empirical Distributions}

 We start by defining the Wasserstein distance between probability measures supported on an arbitrary metric space. Our main reference for classical optimal transport is \cite{villani2009optimal}.

\begin{definition}
  Let $(\Omega,d)$ be a Polish metric space. For $p \geq 1$, let \[\mathcal{P}_p(\Omega) \coloneqq \{\mu: \int_\Omega d(x,x_0)^p d\mu(x) < \infty \ \forall x_0 \in \Omega\}\] denote the set of Borel probability measures on $\Omega$ with finite $p$-th moment. For $\nu, \mu \in \mathcal{P}_p(\Omega)$, define $\Pi(\nu,\mu)$ to be the set of \emph{couplings} of $\nu$ and $\mu$, that is, the set of joint probability measures $\gamma$ on $\Omega \times \Omega$ with marginals $\nu$ and $\mu$, respectively. Then
    \begin{equation}\label{eqn:wasserstein_p_distance}
        W_p(\nu, \mu) \coloneqq \bigg(\inf_{\gamma \in \Pi(\nu, \mu)} \int_{\Omega\times \Omega} d(x,y)^p d\gamma(x,y) \bigg)^\frac{1}{p}
    \end{equation}
    defines a distance on $\mathcal{P}_p(\Omega)$ called the \emph{$p$-Wasserstein distance}. We let $W_2^2(\nu,\mu) \coloneqq (W_2(\nu,\mu))^2$ denote the squared 2-Wasserstein distance. Elements of $\Pi(\nu,\mu)$ are also referred to as \emph{transport plans}, and a $\gamma$ that achieves the infimum of \eqref{eqn:wasserstein_p_distance} is called an \emph{optimal transport plan}.
\end{definition}

For arbitrary $\nu, \mu \in \mathcal{P}_p(\Omega)$, it is not possible to calculate $W_p(\nu,\mu)$ explicitly. However, one can always compute Wasserstein distances between \emph{empirical probability measures}, i.e., discrete probability measures with a finite number of support points. For $n \geq 1$, define the \emph{probability simplex} to be the set
\begin{equation*}
    \Sigma_n \coloneqq \{ u\in \R^n_+: \|u\|_1 = 1 \}.
\end{equation*}
One can represent an empirical probability measure with \emph{weights} $a = (a_1, a_2, ..., a_n) \in \Sigma_n$ and \emph{support points} $X = (x_1, x_2, ..., x_n) \in \Omega^n$ as
$\nu = \sum_{i=1}^n a_i \delta_{x_i}$, 
where $\delta_x$ denotes the Dirac probability measure at $x \in \Omega$. {We let $\mathrm{supp}(\nu)$ denote the \emph{support} of a measure $\nu$. }

In order to calculate Wasserstein distances between empirical probability measures, one can use the solution to a standard form linear program known as the \emph{transportation problem} \cite{bertsimas1997introduction}. For empirical probability measures $\nu = \sum_{i=1}^n a_i \delta_{x_i}$, $\mu = \sum_{j=1}^m b_j \delta_{y_j}$, the set of couplings $\Pi(\nu, \mu)$ can be identified with the set of matrices
    \begin{equation*}
        U(a,b) \coloneqq \{ \gamma \in \R^{n \times m}_+ : \sum_{i=1}^n \gamma_{ij} = b_j, \sum_{j=1}^m \gamma_{ij} = a_i \ \forall i,j\}.
    \end{equation*}
Letting $D_{ij} = d(x_i,y_j)^2$, one can compute the squared 2-Wasserstein distance between $\nu$ and $\mu$ by using linear programming to find a matrix $\gamma$ that solves
    \begin{equation*}
        W_2^2(\nu, \mu) = \min_{\gamma \in U(a,b)} \mathrm{trace}(\gamma^T D).
    \end{equation*}
Throughout the paper, we mildly abuse terminology and conflate couplings (measures) with elements of $U(a,b)$ (matrices), when the meaning is clear from context.

We now recall some useful terminology. Let $(X,\nu)$ and $(Y,\mu)$ be probability spaces, and let $T:X \to Y$ be a measurable map. The \emph{pushforward} of $\nu$ by $T$ is the measure $T_\# \nu$ on $Y$ defined by $(T_\# \nu)(A) = \nu(T^{-1}(A))$. The map $T$ is called \emph{measure-preserving} if $T_\# \nu = \mu$; in the context of optimal transport, such a map is called a \emph{transport map}. Observe that a transport map always leads to a coupling of $\nu$ and $\mu$, via $(\mathrm{id}_X \times T)_\# \nu$, where $\mathrm{id}_X \times T:X \to X \times Y$ is the map $x \mapsto (x,T(x))$. If a transport plan (coupling) $\gamma$ is induced by a transport map in this manner, we say that $\gamma$ is a \emph{deterministic coupling}. This is to emphasize the point that a general transport plan is \emph{probabilistic}, in the sense that mass at a single point in $\mathrm{supp}(\nu)$ can be assigned to multiple points in $\mathrm{supp}(\mu)$, whereas this behavior does not occur when the plan is induced by a map.

\subsection{Wasserstein Barycenters and Fr\'echet Variance}

Using the metric structure of $\mathcal{P}_p(\Omega)$ one can define Fr\'echet means, also referred to as \emph{Wasserstein barycenters} \cite{agueh2011barycenters} in the OT literature.
\begin{definition}\label{def: barycenter}
A \emph{p-Wasserstein barycenter} of measures $\mu_\ell \in \mathcal{P}_p(\Omega)$, $\ell=1,...,N$, is any measure $\nu$ satisfying
\begin{equation}\label{eq: barycenter}
    \nu \in \mathrm{argmin}_{\mu \in \mathcal{P}_p(\Omega)} \left( \sum_{\ell=1}^N {W_p^p}(\mu, \mu_\ell) \right)\ .
\end{equation}
\end{definition}

While there exist methods to calculate Wasserstein barycenters for sets of empirical probability measures \cite{anderes2015discrete}, they are often computationally intractable for large datasets. For the applications presented in Section \ref{sec: numerical experiments}, we use \emph{free support} Wasserstein barycenter algorithms \cite{cuturi2014fast, peyre16} to compute approximations of Wasserstein barycenters for sets of empirical probability measures. {The free support barycenter algorithm (\cite{cuturi2014fast}, Algorithm 2) seeks to minimize Eq.~\eqref{eq: barycenter} with respect to the support points of the barycenter (for some fixed weight vector $a$) by minimizing a local quadratic approximation of the gradient of the transport cost.} Given an estimate of the barycenter $\nu^{t}$ with weights $a$ and locations $X^{t}$, and letting $Y^\ell$ be a matrix containing the locations of a measure $\mu_\ell$ with weights $b^\ell$, $\ell=1,...,N$, with $\gamma_{t} \in U(a,b^\ell)$ an optimal coupling of $\nu^{t}$ and $\mu_\ell$, one uses the update rule \cite{cuturi2014fast}

\begin{equation*}
    X^{t+1} = \frac{1}{N}\sum_{\ell=1}^N \mathrm{diag}(a)^{-1} \gamma^{t} Y^\ell.    
\end{equation*}

\begin{figure}
\begin{center}
\includegraphics[height=2.3in]{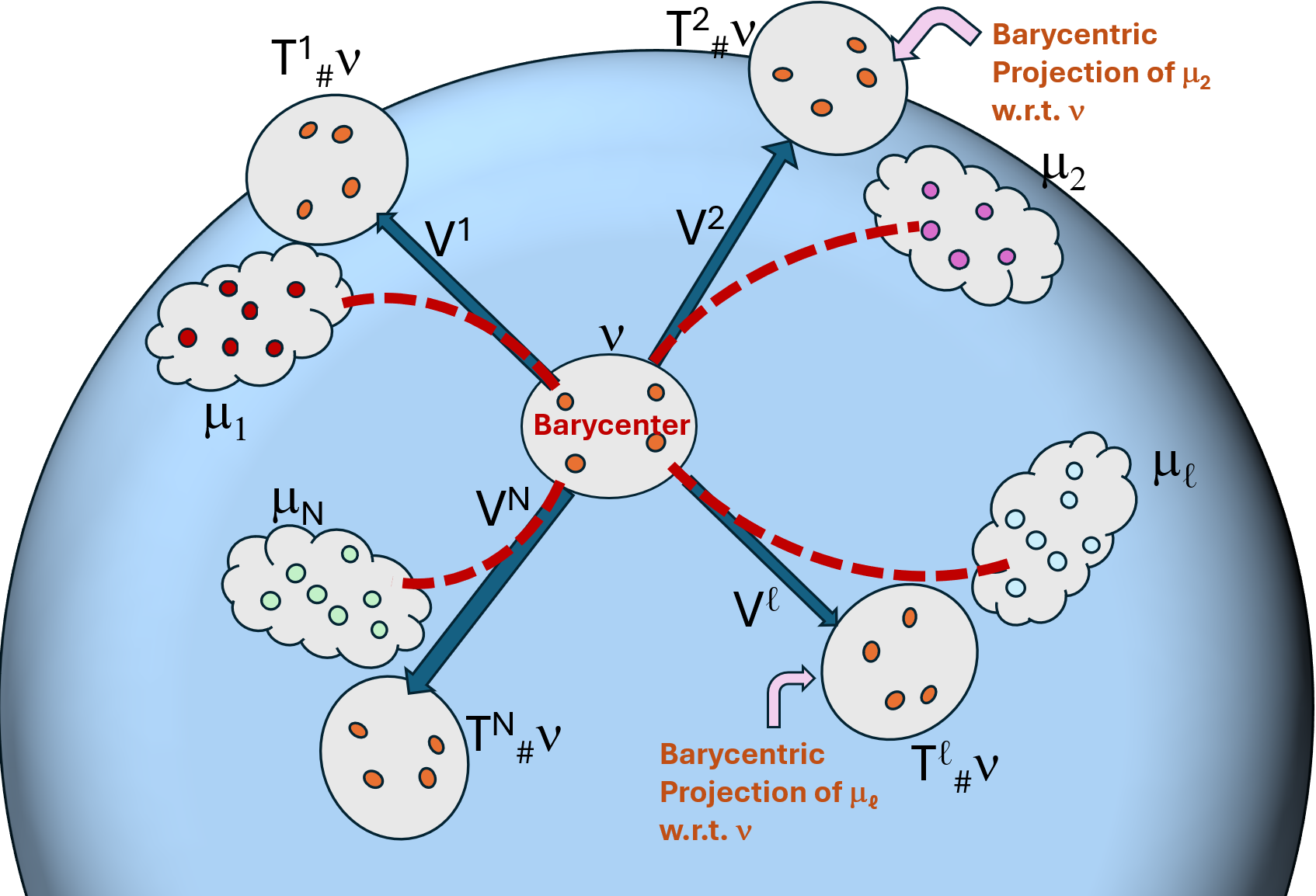}
\end{center}
\caption{Illustration of empirical measures $\mu_1$, $\mu_2$,.., $\mu_l$ and their barycenter $\nu$. The measures $T_\#\nu^l$ denote the barycentric projections of $\nu$ towards $\mu_l$. The vectors $V^l$ represent the Euclidean translations from $\nu$ to $T_\#\nu^l$}  \label{fig:Barycenter-Cartoon}
\end{figure}

Given a set of measures $\mu_\ell$, $\ell=1,...,N$, and a 2-Wasserstein barycenter of those measures $\nu$, the sample \emph{Fr\'echet variance} of the $\mu_\ell$s is given by
\begin{equation*}
    \widehat{Var}_W\coloneqq\frac{1}{N}\sum_{\ell=1}^N W_2^2(\nu, \mu_\ell). 
\end{equation*}
The sample Fr\'echet variance provides a way to quantify the dispersion of data in metric spaces, where standard definitions of variance/covariance cannot necessarily be applied. Figure~\ref{fig:Barycenter-Cartoon} shows a schematic illustration of Wasserstein barycenter of empirical measures $\mu_\ell$, $\ell=1,...,N$. %

\subsection{Linear Optimal Transport}

Many standard statistical methods (e.g., Regression, Principal Components Analysis (PCA), Support Vector Machines (SVM), etc.) are naturally designed for Euclidean data. When working with non-Euclidean data, the typical approach is to \emph{linearize} the data and then apply standard statistical methods. In the Riemannian setting, a standard pipeline used to linearize data is to (1) calculate a Fr\'echet mean of the data, and (2) apply the logarithmic map (calculated with respect to the Fr\'echet mean) to get tangent vector representations of the data, at which point statistical methods for Euclidean data can be applied (for examples, see \cite{srivastava-klassen:2016,guigui2023introduction}). 

Linear Optimal Transport (LOT) \cite{Wang2013} is an analogous pipeline for linearizing data in a Wasserstein space with respect to a template measure (for which a natural choice is the Wasserstein barycenter). In order to linearize measure data, LOT uses \emph{barycentric projection}. 

\begin{definition}\label{def: barycentric projection}(See \cite[Def. 5.4.2]{ambrosio2008gradient})
Let $\nu = \sum_{i=1}^n a_i \delta_{x_i}$, $\mu = \sum_{j=1}^m b_j \delta_{y_j}$ be empirical probability measures on $\R^d$, and let $\gamma$ be an optimal coupling of $\nu$ and $\mu$. Then the \emph{barycentric projection map induced by $\gamma$}, $T:\mathrm{supp}(\nu) \to \R^d$, is defined to be
\begin{equation*}
    T(x_i) \coloneqq \sum_j \frac{\gamma_{ij}}{a_i} y_j, \ i = 1,...,n,
\end{equation*}
and the \emph{barycentric projection of $\mu$ w.r.t. $\nu$} is given by 
\begin{equation}\label{eq: barycentric projection}
    T_\#\nu = \sum_{i=1}^n a_i \delta_{T({x}_i)}.
\end{equation}
\end{definition}

One can use barycentric projection maps to define a vector field on $\mathrm{supp}(\nu)$ as follows. We first identify $\mathrm{supp}(\nu)$ with a matrix $x \in \R^{n \times d}$. Then the vector associated with the $i$th support point of $\nu$ is given by
\begin{equation}
    V_i = T(x_i)-x_i ~{ \in \real^{d},\ \ i=1,2,\dots, n.}
\end{equation}
Importantly, for a template measure $\nu$ with $n$ support points, the vector fields calculated for any $\mu_\ell = \sum_{j=1}^{m_\ell} b_j \delta_{y_j^\ell}$ will be of fixed dimension; i.e., the size of $V^\ell \in \R^{n \times d}$ will not depend on $m_\ell$. Thus, one can vectorize the $V^\ell$s and apply standard statistical methods to these vector representations of the measures. {We refer to the $V^\ell$s as LOT embeddings of the $\mu_\ell$s.}

The fact that Wasserstein spaces are not Riemannian manifolds leads to some issues with the analogy to the pipeline for linearizing data in the Riemannian setting. As noted in \cite{cazelles2017log}, barycentric projection with respect to an empirical probability measure is a lossy process in general. Thus, a natural question is - ``How well do LOT embeddings describe the original data?". {In the next section, we show that, for measures supported on Euclidean spaces, one can quantify exactly the percentage of Fr\'echet variance explained by LOT embeddings for three popular OT distances.}

\section{Fr\'echet Variance Decompositions with LOT}\label{sec: novel results}

{In this section, we present conditions under which Linear Optimal Transport can be used to decompose Fr\'echet variance with respect to the Wasserstein, Gromov-Wasserstein, and Fused Gromov-Wasserstein distances. 

\subsection{Decomposition of Wasserstein Distance} 

We begin with a proposition first presented in \cite{alfonsi2020squared}.}

\begin{proposition}\label{prop: w decomp} Let $\nu = \sum_{i=1}^n a_i \delta_{x_i}$ and $\mu = \sum_{j=1}^m b_j \delta_{y_j}$ be empirical probability measures supported on $\R^d$, let $\gamma \in U(a,b)$ be an optimal coupling of $\nu$ and $\mu$, and let $T$ be the barycentric projection map induced by $\gamma$ {as defined in Eq.~\ref{eq: barycentric projection}}. 
Then
    \begin{equation}\label{eq: decomp}
        W_2^2(\nu, \mu) = W_2^2(\nu, T_\#\nu) +  \sum_{ij} \gamma_{ij} \|T(x_i) - y_j\|^2. 
    \end{equation}
\end{proposition}
Although this result was derived in \cite{alfonsi2020squared}, for the sake of completeness, and due to its correspondence with the proof of Theorem~\ref{thm: gw decomp}, we include a proof of Proposition \ref{prop: w decomp} in Appendix \ref{app: w decomp}. Because the terms on the RHS of Eq.~\eqref{eq: decomp} correspond to the transport cost of a deterministic coupling (of $\nu$ and $T_\# \nu$) and a probabilistic coupling (of $T_\# \nu$ and $\mu$), we refer to these as the \emph{deterministic} and \emph{probabilistic components} of the decomposition, respectively. More on that later. The proof of the proposition shows that one can find a similar decomposition into deterministic and probabilistic components for any coupling $\gamma \in U(a, b)$; however, for non-optimal couplings, the first term in \eqref{eq: decomp} will no longer be a squared Wasserstein distance.

We also provide here an interpretation of the decomposition which is novel, to the best of our knowledge. Given an arbitrary coupling $\gamma \in U(a,b)$, one obtains a collection of $m$ vector fields on $\mathrm{supp}(\nu)$: the vectors $W^1_i,\ldots,W^m_i$ based at the point $x_i$ are given by
\[
W_i^j = \gamma_{ij}\cdot(y_j - x_i).
\]
We say that the coupling $\gamma$ is \emph{deterministic} if there is exactly one nonzero vector based at each $x_i$. On the opposite end of the spectrum, we say that $\gamma$ is \emph{purely probabilistic} if, for each $i$, 
\[
\sum_{j=1}^m W_i^j = 0.
\]
Intuitively, a deterministic coupling acts on the density $\nu$ by moving its points around in the ambient space---in fact, it is clear from the definition that a coupling is deterministic if and only if it is induced by a transport map. On the other hand, a purely probabilistic coupling acts on $\nu$ by completely redistributing its mass; it has no component which acts by moving any point in the ambient space. 
The Eq.~\eqref{eq: decomp} decomposes an optimal coupling $\gamma$ into a deterministic coupling between $\nu$ and $T_\# \nu$ and a purely probabilistic coupling $\gamma$ between $T_\# \nu$ and $\mu$. 
To see the latter point, observe that
\begin{align*}
    \sum_{j=1}^m W_i^j &= \sum_j \gamma_{ij}(y_j - T(x_i)) = \sum_j \gamma_{ij} y_j - T(x_i) \sum_j \gamma_{ij} \\
    &= \sum_j \gamma_{ij} y_j - \left(\sum_j \frac{\gamma_{ij}}{a_i}y_j\right) a_i = 0.
\end{align*}

\subsection{Wasserstein Variance Decomposition}
Note that, since Eq.~\eqref{eq: decomp} applies to a single squared 2-Wasserstein distance, it also works for averages of squared 2-Wasserstein distances. In particular, the sample Fr\'echet variance of a set of measures $\mu_\ell$, $\ell=1,...,N$ calculated with respect to a free support barycenter $\nu^n$ with $n$ support points, denoted $\widehat{Var}_W^n$, can be decomposed into deterministic and probabilistic components. 

\begin{definition}
    Let {$\mu_\ell = \sum_{j=1}^{m_\ell} b^\ell_j \delta_{y^\ell_j}$, ${\ell=1,...,N}$ be empirical probability measures with $\nu^n = \sum_{i=1}^n a_i \delta_{x_i}$} a {free support} barycenter of the $\mu_\ell$s with $n$ support points, $\gamma^\ell$ an optimal coupling of $\nu^n$ and $\mu_\ell$, and $T^\ell$ the barycentric projection map induced by $\gamma^\ell$. Then 

\begin{align}\label{eq: var decomp}
    & \widehat{Var}_W^n\coloneqq\frac{1}{N}\sum_{\ell=1}^N W_2^2(\nu^n,\mu_\ell) \nonumber \\
    & = \frac{1}{N}\sum_{\ell=1}^N W_2^2(\nu^n,T^\ell_\#\nu^n) + \frac{1}{N}\sum_{\ell=1}^N \sum_{ij} \gamma^\ell_{ij} \|T^\ell(x_i) - y_j\|^2 .
\end{align}

\end{definition}
Note that the deterministic component (the first term on the RHS of Eq.~\eqref{eq: var decomp}) corresponds to the sample Fr\'echet variance of the barycentric projections of the $\mu_\ell$'s w.r.t. $\nu^n$, and thus can be used to quantify the percentage of the variance in the $\mu_\ell$s that is explained by the LOT embedding.

It is worth noting that the Law of Total Variance \cite{CaseBerg:01} can be seen as a special case of this decomposition, where $d=1$, $n=1$ and $N\geq2$, or in other words, a `Law of Total Fr\'echet Variance' defines a Pythagorean-like theorem in the 2-Wasserstein space for empirical probability measures supported on $\R^d$ . This decomposition thus also relates to the $F$-statistic used in One-Way Analysis of Variance \cite{fisher1970statistical, box1978statistics}: observe that if the $\mu_\ell \in \mathcal{P}(\R)$ (treated as sample data rather than measures) all have uniform weights, and we take $n=1$, one can calculate an $F$-statistic as

\begin{equation}\label{eq: F}
    F^{n,d} = \frac{(\sum_{\ell=1}^N m_\ell) - N}{N - 1} \frac{\sum_{{\ell^*}=1}^N W_2^2(\nu^n,T^{\ell^*}_\#\nu^n)}{\sum_{\ell=1}^N \sum_{ij} \gamma^\ell_{ij} \|T^\ell(x_i) - y^\ell_j\|^2}.
\end{equation}

This trivially extends to a weighted ANOVA if the $\mu_\ell$s have non-uniform weights. However, this formula can also be used for $n>1$ (testing for equality of $n$-support barycentric projection) and/or $d>1$ (multivariate data). In an intuitive sense, the statistic can be seen as testing for equality of cluster means across data sets; the numerator captures the average `between group' variation of cluster means $T^\ell(x_i)$ about $\frac{1}{N} \sum_{\ell=1}^N T^\ell(x_i)$, while the denominator captures the average `within group' variance of all clusters across all datasets. In this interpretation, standard one-way ANOVA would be a special case where the datasets are all assumed to have 1 cluster.
To be a bit more rigorous, note that $F^{n,d}$ will equal zero if all of the couplings $\gamma^\ell$ are purely probabilistic and will increase without bound as the percent of Fr\`echet Variance explained by the deterministic component increases, being undefined if all of the couplings are deterministic. 

Although the question of how the values $n$ and $d$ affect the distribution of the statistic is quite interesting, deriving the distribution of $F^{n,d}$ for $n>1$ and/or $d>1$ is beyond the scope of this work. Instead, in section \ref{sec: anova}, we present examples of how one can conduct a permutation test for equality of $n$-support barycentric projection using $F^{n,d}$.

In the following sections, we extend the decomposition of Fr\'echet Variance presented in \eqref{eq: var decomp} to two other popular OT distances.

\subsection{Gromov-Wasserstein Variance Decomposition}\label{sec: gromov}

The Gromov-Wasserstein distance~\cite{memoli2007use, memoli2011gromov, chowdhury2019gromov} is an optimal transport-based distance between \emph{measure networks}, as we now recall. 

    \begin{definition} A \emph{measure network} is a triple $(X,\omega_X, \nu)$, where $X$ is a Polish space, $\nu$ is a Borel probability measure on $X$, and $\omega_X$ is a square integrable function on {$X\times X$ (w.r.t $\nu \otimes \nu$)}. The \emph{2-Gromov-Wasserstein distance}  is given by
    \begin{align*}
        & GW_2^2(\mathcal{X},\mathcal{Y}) \\
        & = \inf_{\gamma \in \Pi(\nu, \mu)}\int_{(X\times Y)^2} (\omega_X(x,x') - \omega_Y(y,y'))^2 \\
        &\hspace{2in} \gamma(dx,dy) \gamma(dx',dy') 
    \end{align*}
    \end{definition}
    Gromov-Wasserstein optimal couplings always exist~\cite[Theorem 2.2]{chowdhury2019gromov}, so that the infimum in the definition is actually a minimum. We are particularly interested in finite measure networks, where we assume that $X = \{x_1,\ldots,x_n\}$, $Y = \{y_1,\ldots,y_m\}$ and that the measures are given by $\nu = \sum_{i=1}^n a_i \delta_{x_i}$, $\mu = \sum_{j=1}^m b_j \delta_{y_j}$. We adopt the notation $A_{ik} = \omega_X(x_i,x_k)$ and $B_{jl} = \omega_Y(y_j,y_l)$, so that $GW_2^2(\mathcal{X}, \mathcal{Y})$ can be written as
    \begin{equation*}
        GW_2^2(\mathcal{X},\mathcal{Y}) = \min_{\gamma \in U(a,b)} \sum_{ijkl} \gamma_{ij} \gamma_{kl} |A_{ik} - B_{jl}|^2
    \end{equation*}
    One can define a natural notion of barycentric projection in the GW setting as follows (see~\cite[Definition 1]{nguyen2023linear}).

    \begin{definition}\label{def: GW barycentric projection}
        Let $\mathcal{X} = (X,\omega_X, \nu)$ and $\mathcal{Y} = (Y,\omega_Y, \mu)$ be two finite measure networks. Let $\gamma$ be a Gromov-Wasserstein optimal coupling of $\nu$ and $\mu$, and define 
        \[
        \omega_C(x_i,x_k) \coloneqq \sum_{jl} \frac{\gamma_{ij}}{a_i}\frac{\gamma_{kl}}{a_k}B_{jl}.
        \]
        Then the \emph{GW barycentric projection of $\mathcal{Y}$ with respect to $\mathcal{X}$} is given by
        \begin{equation*}
            \mathcal{T}\coloneqq (X, \omega_C, \nu).
        \end{equation*}
    \end{definition}
    
    Using this notion of barycentric projection, one can extend the decomposition in Eq.~\eqref{eq: decomp} to the Gromov-Wasserstein setting.
    
    \begin{theorem}\label{thm: gw decomp}
        Let $\mathcal{X} = (X,\omega_X, \nu)$ and $\mathcal{Y} = (Y,\omega_Y, \mu)$ be two finite measure networks, $\nu = \sum_{i=1}^n a_i \delta_{x_i}$, and $\mu = \sum_{j=1}^m b_j \delta_{y_j}$. Let $\gamma$ be a Gromov-Wasserstein optimal coupling of $\mathcal{X}$ and $\mathcal{Y}$, define $B_{jl} = \omega_Y(y_j,y_l)$ and $C_{ik} = \sum_{jl} \frac{\gamma_{ij}}{a_i}\frac{\gamma_{kl}}{a_k}B_{jl}$, and let $\mathcal{T}$ denote the GW barycentric projection of $\mathcal{Y}$ with respect to $\mathcal{X}$. Then
    \begin{align}\label{eq: GW decomp}
        & GW_2^2(\mathcal{X},\mathcal{Y}) = GW_2^2(\mathcal{X},\mathcal{T}) + \sum_{ijkl} \gamma_{ij} \gamma_{kl} |C_{ik} - B_{jl}|^2 \nonumber \\
        & = GW_2^2(\mathcal{X},\mathcal{T}) + (\mathrm{diam}_2(\mathcal{Y})^2 - \mathrm{diam}_2(\mathcal{T})^2),
    \end{align}
    where 
    \begin{equation}\label{eqn:2-diameter}
    \mathrm{diam}_2(\mathcal{X}) \coloneqq \left(\iint_{X \times X} \omega_X(x,x')^2 \nu(dx)\nu(dx') \right)^{1/2}.
    \end{equation}
    \end{theorem}
    Note that $C_{ik}$ is the evaluation of $\omega_C$ on the pairs $(x_i, x_k)$.
    
    The \emph{2-diameter} of a measure network, as defined in \eqref{eqn:2-diameter}, gives a natural notion of its size, and appears in certain estimates of GW distances (see~\cite{memoli2011gromov,chowdhury2019gromov}). The proof of the theorem is provided in Appendix \ref{app: gw decomp}. The proof shows that an optimal coupling of $\mathcal{X}$ and $\mathcal{T}$ is given by the identity coupling $(\mathrm{id}_X \times \mathrm{id}_X)_\#\nu$.
    Additionally, it shows that \emph{any} coupling $\gamma$ separates into deterministic and probabilistic components, though, again, the first term will only necessarily be $GW_2^2$ if $\gamma$ is optimal. 

    \subsection{Fused Gromov-Wasserstein Variance Decomposition}\label{sec: fused}

    The Fused Gromov-Wasserstein distance \cite{vayer2019optimal} is an optimal transport distance between `structured data objects' - measure networks whose nodes are elements of the same metric space. For the purposes of defining a variance decomposition, we are only interested in the special case of the Fused 2-Gromov-Wasserstein distance. 

    \begin{figure*}
    \centering
    \includegraphics[width=\textwidth]{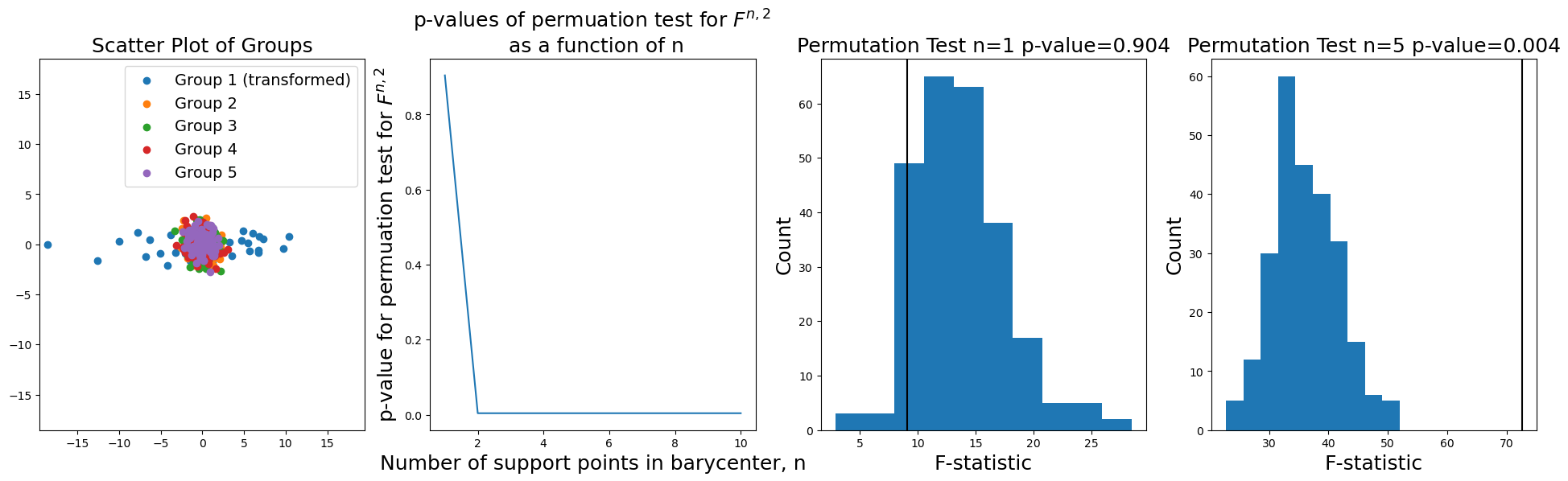}
    \caption{Left: Sample data from 5 groups; Middle Left: p-values for permutations test for different values of n; Middle Right: Permutation distribution and observed F statistic for n=1; Right: Permutation distribution and observed F statistic for n=5.}
    \label{fig:anova}
\end{figure*}

    \begin{definition}
    Let $(\Omega,d)$ be a metric space, $\mathcal{X} = (X, \omega_X, \nu)$ and $\mathcal{Y} = (Y, \omega_Y, \mu)$ be two  measure networks with $X,Y \subset\Omega$, and $\alpha \in [0,1]$. Then the \emph{squared $\alpha$-Fused 2-Gromov-Wasserstein distance} is given by 
        \begin{align*}
            & FGW_{2,\alpha}^2(\mathcal{X}, \mathcal{Y}) = \inf_{\gamma \in \Pi(\nu, \mu)}\int_{X\times Y \times X \times Y} [\alpha d(x,y)^2 \\
            & + (1-\alpha)(\omega_X(x,x') - \omega_Y(y,y'))^2] \ \gamma(dx,dy) \gamma(dx',dy').
        \end{align*}
    \end{definition}
    Using Definitions \ref{def: barycentric projection} and \ref{def: GW barycentric projection}, it is straightforward to define a notion of barycentric projection in the FGW setting, for measure networks supported on $\R^d$ (see~\cite[Definition 1]{nguyen2023linear}).

    \begin{definition}\label{def: FGW barycentric projection}
        Let $\mathcal{X} = (X,\omega_X, \nu)$ and $\mathcal{Y} = (Y, \omega_Y, \mu)$ be two finite measure networks with $X = \{x_i \in \R^d, i = 1,...,n\}$, $Y = \{y_j \in \R^d, j = 1,...,m\}$, $\nu = \sum_{i=1}^n a_i \delta_{x_i}$, and $\mu = \sum_{j=1}^m b_j \delta_{y_j}$. Let $\gamma$ be an $\alpha$-Fused 2-Gromov-Wasserstein  optimal coupling of $\mathcal{X}$ and $\mathcal{Y}$, define $T(x_i) = \sum_j \frac{\gamma_{ij}}{a_i}y_j$, $T(X) \in \R^{n \times d}$ such that $T(X)_i = T(x_i)$, $B_{jl} = \omega_Y(y_j,y_l)$ and $\omega_C(x_i,x_k) = \sum_{jl} \frac{\gamma_{ij}}{a_i}\frac{\gamma_{kl}}{a_k}B_{jl}$. Then the FGW barycentric projection of $\mathcal{Y}$ with respect to $\mathcal{X}$ is given by
        \begin{equation*}
            \mathcal{T} := (T(X), \omega_C, T_\#\nu)
        \end{equation*}
    \end{definition}
    Given equations \eqref{eq: decomp} and \eqref{eq: GW decomp}, it is easy to see that one can define a decomposition of the squared 2-FGW distance for structured data objects supported on $\R^d$ as well. For notational simplicity, we define transport cost functions as follows.

    \begin{definition}\label{def: transport cost} Let $\mathcal{X} = (X, \omega_X, \nu)$ and $\mathcal{Y} = (Y, \omega_Y, \mu)$ be two measure networks with $X,Y \subset\R^d$, $\nu = \sum_{i=1}^n a_i \delta_{x_i}$ and $\mu = \sum_{j=1}^m b_j \delta_{y_j}$. Let $\gamma$ be any coupling of $\nu$ and $\mu$. Then the \emph{squared 2-Wasserstein transport cost} of $\gamma \in \Pi(\nu,\mu)$ is defined to be
    \begin{equation*}
        C_W(\gamma) \coloneqq \sum_{ij} \gamma_{ij} \|x_i - y_j\|^2 
    \end{equation*}
    the \emph{squared 2-Gromov-Wasserstein transport cost} of $\gamma$ is defined to be
    \begin{equation*}
        C_{GW}(\gamma) \coloneqq \sum_{ijkl} \gamma_{ij} \gamma_{kl} (\omega_X(x_i, x_k) - \omega_Y(y_j, y_l))^2
    \end{equation*}
    and the \emph{squared $\alpha$-Fused 2-Gromov-Wasserstein transport cost} of $\gamma$ is defined to be
    \begin{equation*}
        C_{FGW}^\alpha(\gamma) \coloneqq \alpha C_W(\gamma) + (1- \alpha)C_{GW}(\gamma)
    \end{equation*}
    \end{definition}
    Due to the fact that the separability of the decompositions given in Eq.~\eqref{eq: decomp} and Eq.~\eqref{eq: GW decomp} do not depend on the optimality of $\gamma$ (see the comments following each result), we get the following corollary. 
    
    \begin{corollary}\label{cor: FGW decomp}
    Let $\mathcal{X} = (X,\omega_X, \nu)$ and $\mathcal{Y} = (Y,\omega_Y, \mu)$ be two measure networks, as in Definition \ref{def: FGW barycentric projection}. Let $\alpha \in [0,1]$ and let $\gamma$ be an $\alpha$-Fused 2-Gromov-Wasserstein optimal coupling of $\nu$ and $\mu$, $T(x_i) = \sum_j \frac{\gamma_{ij}}{a_i}y_j$, and $\pi = (T, \mathrm{id}_Y)_\#\gamma$, where $(T,\mathrm{id}_Y):X \times Y \to T(X) \times Y$ is the map $(x,y) \mapsto (T(x),y)$. Then
    \begin{equation*}
        FGW_{2,\alpha}^2(\mathcal{X}, \mathcal{Y}) = FGW^2_{2,\alpha}(\mathcal{X}, \mathcal{T}) +  C^{\alpha}_{FGW}(\pi) 
    \end{equation*}

    \begin{proof}
     We verify this by direct calculation, using the Fr\'{e}chet variance decompositions Proposition \ref{prop: w decomp} and Theorem \ref{thm: gw decomp}:
        \begin{align*}
            & FGW_{2,\alpha}^2(\mathcal{X}, \mathcal{Y}) = \alpha C_{W}(\gamma) + (1-\alpha) C_{GW}(\gamma)\\
            & = \alpha (C_{W}(T) + C_{W}(\pi)) + (1-\alpha)(C_{GW}(T) + C_{GW}(\pi)))\\
            & =  C^{\alpha}_{FGW}(T) +  C^{\alpha}_{FGW}(\pi)\\
            & = FGW^2_{2,\alpha}(\mathcal{X}, \mathcal{T}) +  C^{\alpha}_{FGW}(\pi),
        \end{align*}
        where the last equality holds by \cite[Lemma 1, part 1]{nguyen2023linear}.
    \end{proof}
        
        Thus, given a data set of structured data objects $\mu_\ell, \ell=1,...,N$, the sample $\alpha$-Fused 2-Gromov-Wasserstein Fr\'echet variance can be decomposed as

        \begin{align*}
            &\widehat{Var}^n_{FGW_{2,\alpha}} \coloneqq \frac{1}{N} \sum_{\ell=1}^N FGW_{2,\alpha}^2(\mathcal{X},\mathcal{Y}^\ell) \\
            & = \frac{\alpha}{N} \sum_{\ell=1}^N FGW^2_{2,\alpha}(\mathcal{X}, \mathcal{T}^\ell) + \frac{1-\alpha}{N} \sum_{\ell=1}^N C^{\alpha}_{FGW}(\pi^\ell)
        \end{align*}
    \end{corollary}

\section{Applications to Data Analysis}\label{sec: numerical experiments}

In this section, we demonstrate the theoretical tools developed in section \ref{sec: novel results} to represent and analyze datasets of measures and measure networks.

\subsection{Illustrating $F^{n,d}$ Using Simulated Data}\label{sec: anova}

Here we present results for permutation tests of equality of $n$-support barycentric projection, for $d=2$. Specifically, we generate sample data (with sample size $m=100$) from standard Gaussians (on $\R^2$) for $N=5$ groups, and apply the transformation $T(x,y) = (x^3,y)$ to one of the Gaussian samples. The left panel of Figure \ref{fig:anova} presents a scatter plot of the data.  We then calculate the test statistic $F^{n,2}$ for this data, and estimate the permutation distribution of $F^{n,2}$ with k=250 permutations, for $n = 1,...,10$. Importantly, this transformation does not change the expected value of the mean of the transformed data (i.e. $E[T(X,Y)] = E[(X,Y)]= {\bf 0}$), and so the expected value of the 1-support barycenter of the sample should be unchanged. The second panel from the left presents the $p$-values of a permutation test of the $F^{n,2}$ statistic on the data, for different values of $n$. As expected, the test fails to reject the hypothesis of 1-support barycentric projection, but rejects the hypothesis of equality of $n$-support barycentric projection for $n>1$. The third and fourth panels from the left present the estimated permutation distributions of $F^{1,2}$ and $F^{5,2}$ calculated on the data.

\subsection{Decompositions}

In the following sections, we study the behavior of the decompositions presented in Section \ref{sec: novel results}, as well as the accuracy of machine learning classifiers applied to LOT embedded data, as a function of the number of support points in a Free Support barycenter. In section \ref{applications: mnist}, we present results for digit classification on the MNIST handwritten digits dataset, in section \ref{applications: imdb} we present results for sentiment analysis on the IMDB-50000 dataset, and in section \ref{applications: dtmri}, we present results on gender classification on DTMRI data from the HCP-YA dataset.

\subsection{MNIST Data}\label{applications: mnist}

\begin{figure*}[t]
    \centering
    \includegraphics[width=\textwidth]{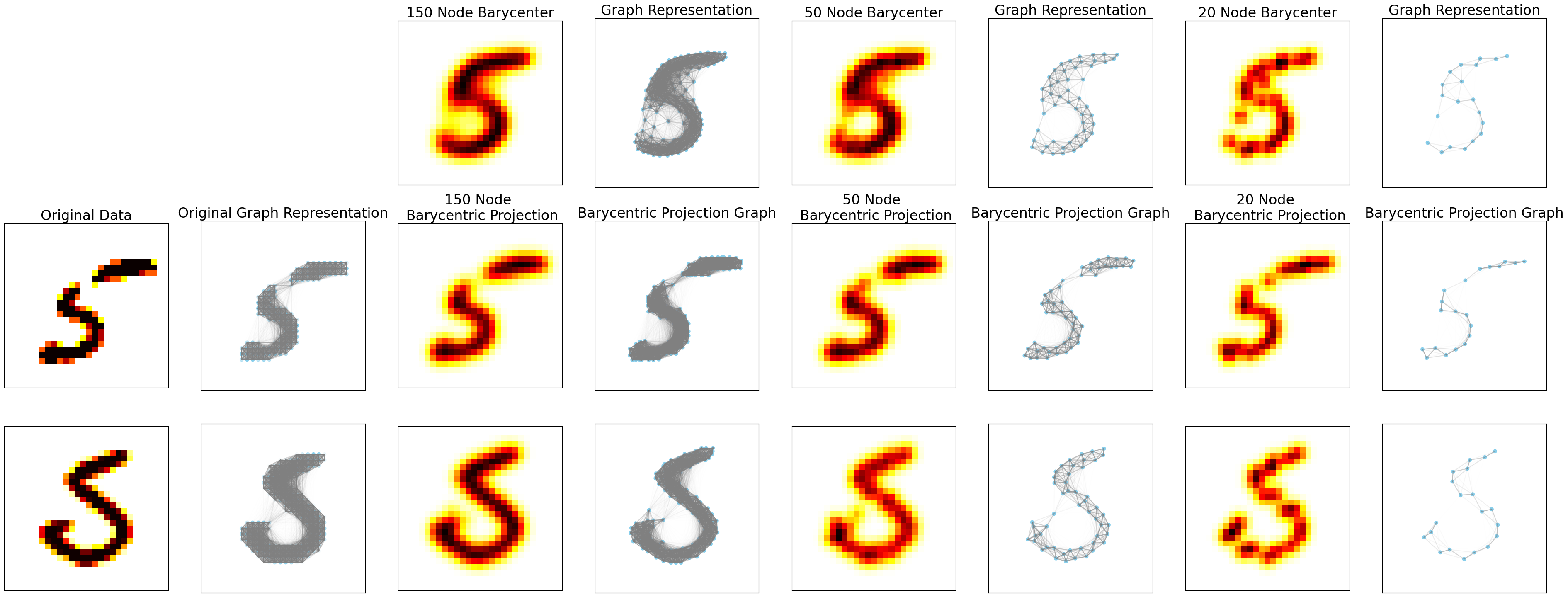}
     \caption{Top Row: Gaussian Kernel Reconstructions and graph representations for free support Fused Gromov-Wasserstein barycenters ($\alpha = 0.5$) of MNIST 5 digit, with n = 150, 50, and 20 nodes; Bottom Rows: Original Data, Gaussian Kernel Reconstructions of Barycentric projections, and graph representations for two 5 digits from the MNIST data set.}
    \label{fig: fgw mnist 1}
\end{figure*}

We begin with experiments studying the variance decomposition using the MNIST handwritten digits dataset. The dataset is a set of grayscale images of handwritten digits, with classes $0,...,9$. We represent each image as an empirical measure network in $\real^2$, using pixel locations as support points/nodes, Euclidean distances between support points as edge weights, and pixel intensities (normalized to sum to 1) as weights. {Using these representations, we use the free support Wasserstein barycenter algorithms \cite{cuturi2014fast, peyre16} to calculate barycenters with different numbers of support points, and calculate LOT embeddings with respect to these barycenters, for $\alpha=0,0.25, 0.5, 0.75, 1$. These LOT embeddings have dimensions that depends on the number of support points/nodes in the free support barycenter, and this allows us to choose the dimension of the LOT embedded data.} 

The top row of Fig.~\ref{fig: fgw mnist 1} presents Gaussian kernel reconstructions (reconstructed using the algorithm described in appendix \ref{app: gaussian kernel}) of free support barycenters and their nodes, for the digit ``5", for $n = 150, 50, 20$ and $\alpha = 0.5$. The bottom two rows of Fig.~\ref{fig: fgw mnist 1} present the original images, and Gaussian kernel reconstructions of the barycentric projections (and the node locations) for two particular observations from the dataset, calculated with respect to the barycenter in the top row of the corresponding column. This plot motivates the use of free support barycenters with reduced support sizes in the LOT pipeline. For example, we see little degradation of the reconstructions going from a $150$-point barycenter to a $50$-point barycenter. While there is noticeably more degradation when we use a $20$-support barycenter, the barycentric projections still appear to maintain their fundamental structure.

\begin{figure*}[t]%[bp]
    \centering
    \includegraphics[width = \textwidth]{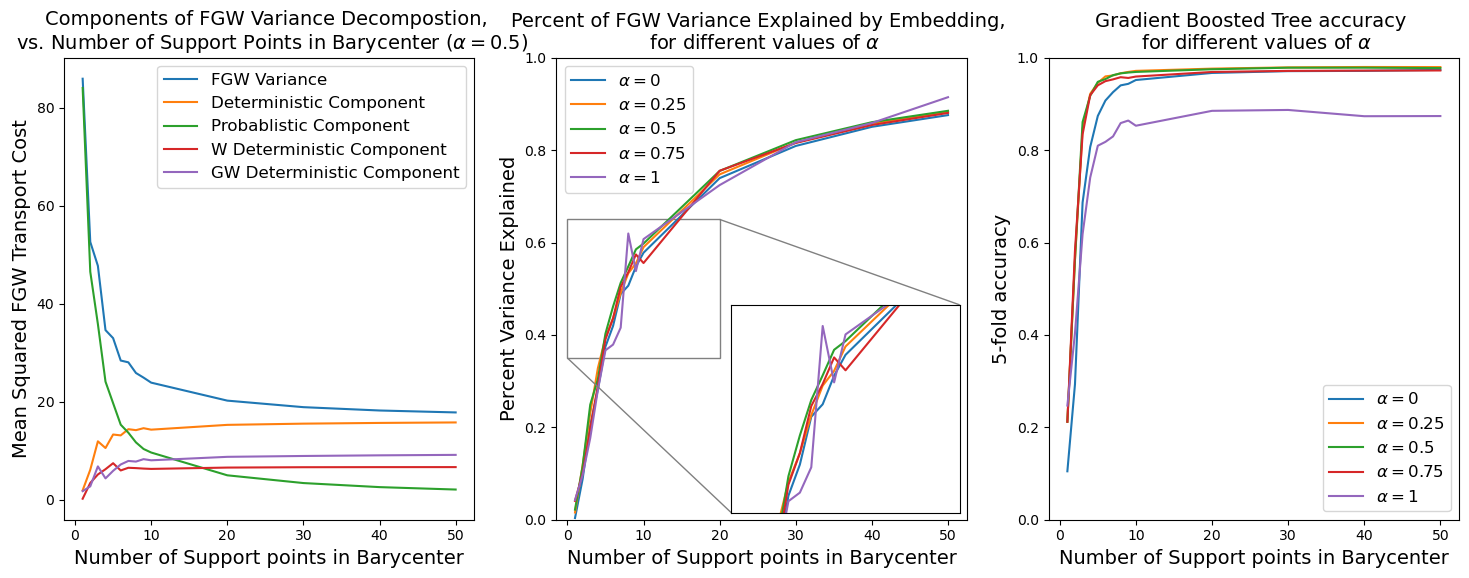}
    \caption{Left: Components of decomposition calculated with respect to a free support Barycenter with different numbers of support points, for MNIST data; Middle: percentage of 2-Wasserstein variance explained by LOT embeddings with respect to a free support Barycenter with different numbers of support points, for MNIST data; Right: Multiclass classification accuracy of a gradient boosted tree built on LOT embeddings with respect to a free support Barycenter with different numbers of support points}
    \label{fig:MNIST 2}
\end{figure*}

Fig.~\ref{fig:MNIST 2} presents a more quantitative exploration of the embedding, based on the decomposition in Corollary~\ref{cor: FGW decomp}, for different values of $n$ and $\alpha$. Specifically, we calculate a free support barycenter on a training set of $5000$ randomly selected digits for $n = 1,...,50$, and then calculate and plot the components of the decomposition, as functions of $n$, on a (seperate) test set of $20000$ randomly selected digits. The leftmost panel presents these results; the $x$-axis corresponds to $n$, the number of support points in the free support barycenter, and the $y$-axis corresponds to the different components of the decomposition. A consequence of Corollary 1 is that one can quantify not
just the percentage of variance explained by LOT embeddings, but also the percentage of variance explained
by the node and edge components of the embedding separately. The left panel of Figure 6 presents these components
of the decomposition as a function of number of support points, for $\alpha$ = 0.5. The middle panel of Figure 6 presents the percentage of variance explained by the deterministic component, for $\alpha$ = 0,0.25,0.5,0.75,1. Importantly, for $\alpha$ = 0,1, we use the Wasserstein and GW barycentric projections - though one could technically use an FGW barycentric projection (i.e., one could still calculate the edge component of the barycentric projection for $\alpha$ = 0 or the node location component for $\alpha$ = 1). Interestingly, the percentage of variance explained is nearly identical for all values of $\alpha$. Considering the left panel of Figure \ref{fig:MNIST 2}, we observe that $\widehat{Var}_{FGW_{2,\alpha}}^n$ decreases as the number of support points increases. This makes sense; more support points should give you more degrees of freedom with which to describe variation, thus resulting in shorter geodesics. The deterministic components, on the other hand, tends to increase with the number of support points. This also makes sense, and mirrors the previous observation; more support points in the barycenter means more of the variation can be described with transport maps. 

The middle panel presents an elbow plot of the percentage of $\widehat{Var}_{FGW, \alpha}^n$ described by the deterministic component; it is the pointwise ratio of the orange curve to the blue curve presented in the left panel. First, note that the percentage of variance explained generally increases with $n$, but it's not strictly non-decreasing (in particular, note the spike at $n=8,9$ for the percentage of variance explained for higher values of $\alpha$ (purple and red curves)). This is likely due (at least in part) to the fact that the barycenters are only approximations, though it is an open question whether the deterministic component should be strictly non-decreasing in $n$. We also note that barycentric projections describe more than $87.5\%$ of $\widehat{Var}_{FGW,0}^{50}$, with other values of $\alpha$ slightly higher. Somewhat surprisingly, the percentage of variance explained is almost identical for all of the distances.

In the right panel, we present the results of an experiment studying the average 5-fold cross validation classification accuracy of a multiclass LightGBM \cite{ke2017lightgbm} gradient boosted tree trained on LOT embeddings (with respect to a free support barycenter built on the training set of $5000$ digits) of $20000$ digits from the MNIST dataset for different values of $n$ and $\alpha$. In order to linearize our measure networks, we vectorize FGW barycentric projections using the mapping

\[ (T(X), \|\cdot\|_2,C) \to
\begin{bmatrix}
{Vec}(T(X)) \\
{Vec}(\text{triu}(2C - \mathrm{diag}(C)))
\end{bmatrix}
\]
where $T(X)$ and $C$ are defined as in Definition~\ref{def: FGW barycentric projection}). Note, for $\alpha = 0$ or $1$, we only vectorize the nodes or edges, respectively.

We find that a gradient boosted tree trained on LOT embeddings with $\alpha = 0.5$, calculated with respect to a free support barycenter with $7$ support points achieves an average of $96.23\%$ test set accuracy, while the performance for 50 support points achieves only a slight improvement of $97.76\%$.  This is a promising result, suggesting that LOT embeddings with respect to a barycenter with a low number of support points can explain a large proportion of information relevant to classification in some datasets. 

Despite the fact that there appears to be little difference between the FGW and Wasserstein decompositions in terms of percentage of variance explained, we find that FGW LOT embeddings outperform Wasserstein LOT embeddings in terms of classification accuracy for lower numbers of support points. However, this comparison is a bit misleading, since the dimension of the embeddings depends on $\alpha$. For a template measure network with $n$ nodes on $\R^d$, the dimension of the linearizations is given by $I(\alpha<1) nd + I(\alpha>0) (n + n(n-1)/2)$. 

Thus, to compare performance in terms of the dimension of the embedding rather than the number of support points, one would have to compare, for example, the $\alpha=0.5, n=7$ embedding to the $\alpha = 0, n =21$ embedding. In this case, the 5-fold accuracy for the $\alpha=0, n=21$ embeddings is $97.09\%$, which is slightly better than the $\alpha=0.5, n=7$ embeddings, which achieved just $96.61\%$. This suggests that LOT embeddings with $\alpha=0$ can be more efficient in terms of dimension than embeddings with $\alpha>0$.       

\subsection{IMDB-50000 Data}\label{applications: imdb}

\begin{figure*}[t]%[bp]
    \centering
    \includegraphics[width = \textwidth]{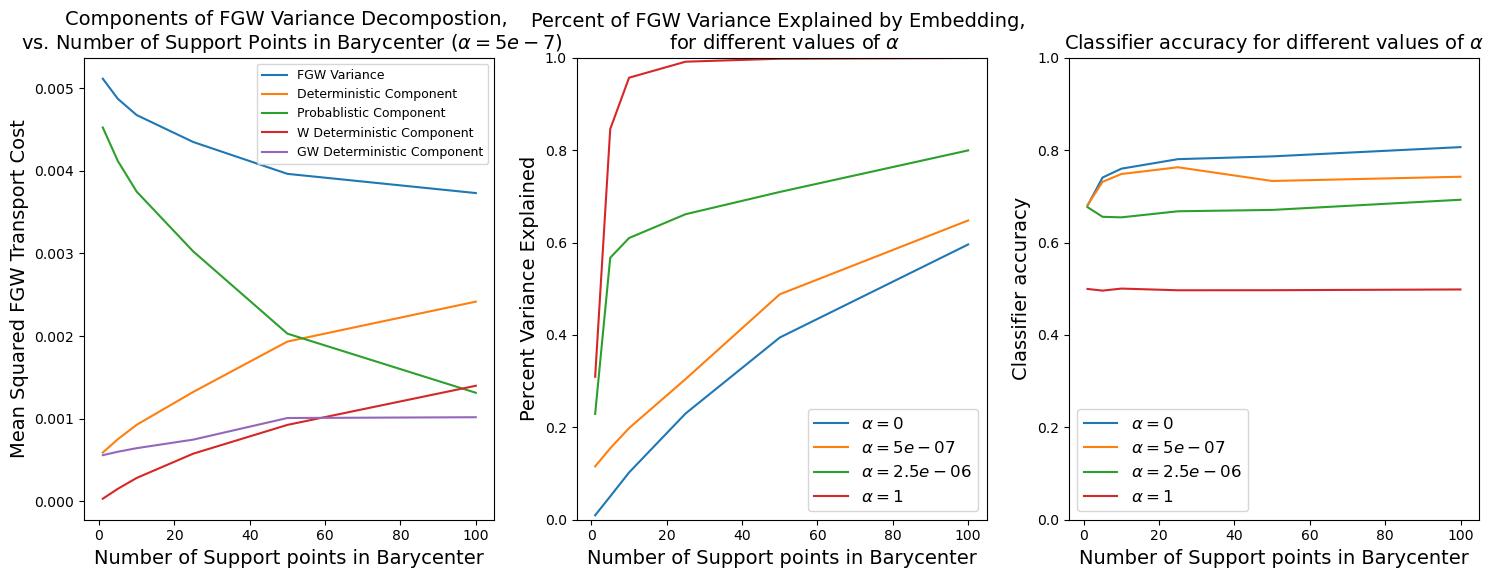}
    \caption{Left: Components of decomposition calculated with respect to a free support Barycenter with different numbers of support points, for IMDB data; Middle: percentage of 2-Wasserstein variance explained by LOT embeddings with respect to a free support Barycenter with different numbers of support points, for IMDB data; Right: Classification accuracy of a gradient boosted tree built on LOT embeddings with respect to a free support Barycenter with different numbers of support points}
    \label{fig: imdb}
\end{figure*}

{Here, we explore an application of LOT to the task of sentiment analysis in the IMDB-50000 dataset. The dataset is made up of 50000 movie reviews (paragraphs of text, ranging from 10-1200 words) from the International Movie Database website, each associated with either a positive or negative sentiment. After training a Word2Vec embedding on the full dataset (with embedding dimension 100), we represent each review as a point cloud in the embedding space. Treating these points as the nodes of a measure network allows us to apply LOT to the data. Interestingly, in this context, LOT can be seen as a sort of linear document embedding of the reviews. Specifically, we represent each review as an empirical measure network $(X,B,\mu)$, where $X$ contains the locations of the vector embeddings of the words in the review, $B_{jl} = |l-j|$ captures the number of words separating $X_j$ and $X_l$ in the review (note, the indices reflect the order of the words in the review, i.e., $y^\ell_j$ corresponds to the vector embedding of the $j$th word in review $\ell$),
and $\mu$ corresponds to a uniform measure supported on $X$. Similar to the MNIST experiment, we calculate barycenters on a training set of $5000$ randomly selected reviews, and then calculate the variance decomposition and conduct 5-fold cross validation with a gradient boosted tree classifier on a test set of $20000$ randomly selected reviews.}

{Figure \ref{fig: imdb} contains the results of the experiment. Specifically, we calculate the variance decomposition and build a gradient boosted tree on LOT embeddings with respect to barycenters with $n=1,5,10,25,50,100$ support points, for $\alpha = 0, 0.0000005, 0.0000025, 1$. These values of $\alpha$ were chosen to balance the percent variance explained by the node versus edge components. 
The middle panel in Figure \ref{fig: imdb} shows the percentage of variance explained by the LOT embedding for different values of $\alpha$. We see that the two lower values of $\alpha$ (which weigh the Wasserstein component more heavily) show much slower growth in percent variance explained than the two higher values. This comes down to 1) the difference in scale between the node and edge components of our measure networks and 2) the fact that the edge matrices all have fairly similar structure in terms of the GW distance, leading to little GW variance in the data for our choice of edge weights.}

{The rationale behind our choice of edge weights was that the graph components of the barycentric projections might capture information about consistent sequences of words across reviews, that could not be captured in the `bag-of-words' information contained in just the node locations. Our classification results, presented in right panel of Figure \ref{fig: imdb}, suggest that this is not the case, with models built only on node locations ($\alpha = 0$) outperforming models for all other values of $\alpha$. Somewhat surprisingly, the 5-fold cross validation accuracy of a model built on LOT embeddings with respect to just 1 support point (corresponding to simply averaging all the points in the point cloud for $\alpha=0$) achieves greater than $68\%$ accuracy. However, we also find that there is an increase in classification accuracy for higher numbers of support points, for $\alpha<1$. The $\alpha=0, n = 100$ LOT embedding performs the best, achieving just over $80\%$ accuracy. Note that the dimension of this embedding is rather high at $nd = 10000$. These results suggest that our edge matrices do not contribute much to the classification accuracy; even when we select an $\alpha$ that balances the W and GW components, we see no benefit in using structured data object representations over simple point clouds - though other edge weights might produce better results. In any case, our results do illustrate that LOT can be used to effectively embed and classify natural language data.}

\subsection{Diffusion Tensor MRI - Human Connectome Project }\label{applications: dtmri}

DTMRI is an approach to MRI that measures the diffusion of water molecules in tissue, such as white matter fiber tracts in the brain. The most common data representations of DTMRI images of white matter fiber tracts are based on `tractography' methods, which seek to reconstruct white matter fiber tracts as sets of curves in $\R^3$. Tractography datasets can be quite large, and only implicitly contain diffusion information in the tangent directions of the individual curves.

Tract density images (TDI), first proposed in \cite{calamante2013super}, offer another data representation for modeling white matter fiber tracts. Instead of representing the tract as a set of 3-d curves, one counts the number of curves passing through each voxel in the image, and represents the tract by associating with each voxel the number of curves that pass through it - effectively representing the tract as an empirical measure on $\R^3$. As the number of curves estimated is often much larger than the number of voxels they pass through, these TDI representations can produce datasets with smaller memory requirements than those produced by tractography methods. Optimal Transport methods for both representations have previously been explored in \cite{feydy2019fast}.

An additional benefit of TDI representations is that they offer a natural way to explicitly include diffusion information in their data representations. Different approaches have been developed to add diffusion information to TDIs, such as `directionally encoded' TDIs \cite{pajevic1999color}, or `Tract Orientation Density Images' (TODI) \cite{dhollander2014track}. The approach taken here is to associate with each voxel in a TDI a diffiusion tensor, i.e., a covariance matrix constructed from the main fiber directions and fractional anisotropy values (as determined by DSI studio \cite{yehdsi}) associated with that voxel. We refer to these representations as Covariance TDIs (CTDI). While these CTDIs will contain less information about diffusivity than TODIs, a benefit is that we are able to represent tracts as empirical probability measures on the space $\R^3 \times \Sym_3^+$ (where $\Sym_3^+$ corresponds to the set of symmetric positive definite matrices), which allows us to run statistical analyses via LOT. While one could run a similar experiment comparing models built on LOT embeddings with different values of $\alpha$, because the DTMRI target measures often have 1000's of support points, this ends up being very computationally expensive. Thus, we choose to study only the Wasserstein LOT embeddings. 

\begin{figure*}
    \centering
    \includegraphics[width=\textwidth]{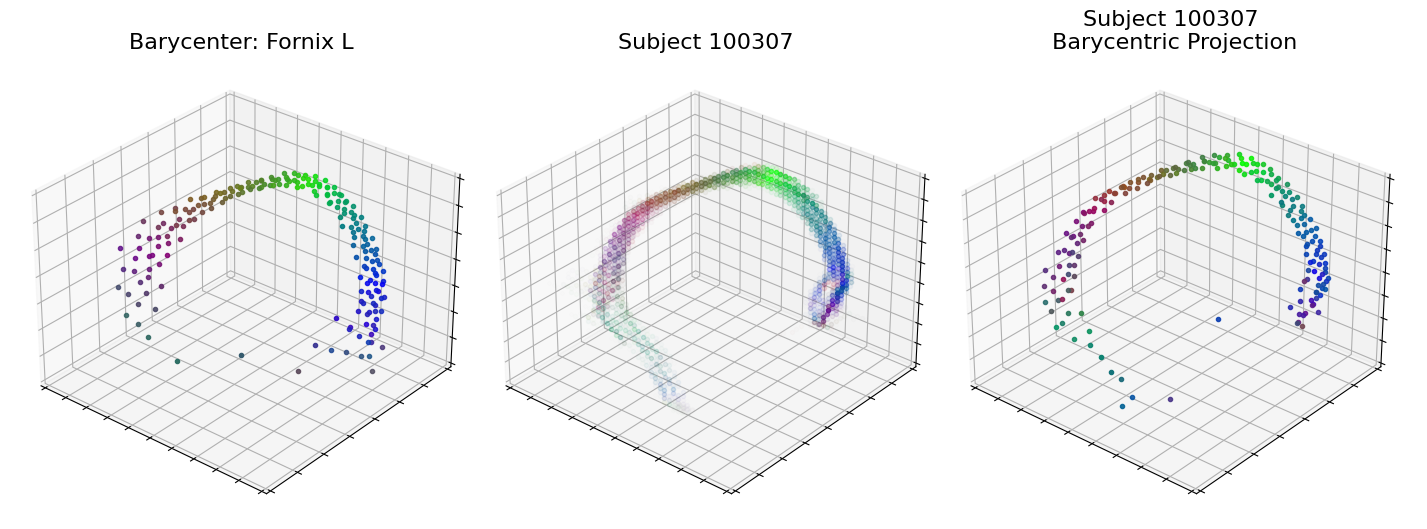}
    \caption{Points represent locations in $\R^3$, $(r,g,b)$ values correspond to (normalized) diagonal elements of covariance matrices. Left: free support barycenter of 'Fornix L' in HCP data (200 support points); Middle: Subject 100307's 'Fornix L' in the HCP data (6980 support points); Right: Barycentric Projection of Subject 100307's 'Fornix L' with respect to calculated barycenter (200 support points).}
    \label{fig: DTMRI 1}
\end{figure*}

Specifically, we apply our framework to the task of classifying gender in the Human Connectome Project - Young Adult (HCP-YA) dataset. The HCP-YA data set is composed of 1200 young adults (aged 22-35) from 300 families. This means the dataset contains many siblings and twins, including identical twins. Clearly, if subjects $i$ and $j$ are identical twins, then $P(gender_i = M | gender_j = M) \approx 1$. However, by removing 1 of each pair of identical twins from the dataset, we believe it is reasonable to assume that $P(gender_i = M | gender_j = M, Family_j = Family_i) \approx .5 \ \forall i, j$ satisfying $Family_j = Family_i$ (and $i \neq j$), i.e. that the gender of one sibling is independent of the gender of their other siblings. While one might expect dependence of CTDIs between siblings, given the assumption that the gender of siblings is independent, we expect any differences learned by our model to be due to differences in CTDIs alone. Thus, while our results don't necessarily represent the performance of the model on a true i.i.d. sample from the population, the classification accuracy should still depend on our model's ability to measure differences in the CTDIs (and hopefully, therefore, in the underlying white matter microstructure).

\subsubsection{Data Representation}

Fiber orientation maps for 1065 subjects from the Human Connectome Project (produced using the pipeline presented in \cite{yeh2016quantifying}) were obtained from the DSI Studio website\footnote{{https://brain.labsolver.org/hcp\_ya.html, FIB 1.25-mm MNI space}}. After preprocessing the data (including removing one of each pair of identical twins), we were left with a dataset with 908 subjects. 
TDIs for each region of interest (ROI) of each subject were produced using DSI studio, and converted to empirical probability measures as follows; given the TDI of a particular ROI for subject $\ell$, we represent it as an empirical measure on $\R^3$
\begin{equation*}
    \tilde{\mu_\ell} = \sum_{j=1}^{m_{\ell}} w_j \delta_{x_j},
\end{equation*}
where $x \in \R^{m_{\ell} \times 3}$ represents the voxel locations that make up the ROI of subject $\ell$ (as determined by DSI Studio), and $w_j$ denotes the number of curves present in the voxel at $x_j$. We then produce a covariance matrix $\Sigma_j$ for each $x_j$, using the fiber orientations $u_{x_j}^r \in \R^{3 \times 1}$ and fractional anisotropy $\mathsf{FA}_{x_j}^r \in \R^+$ (obtained from the fiber orientation map) associated with that voxel

\begin{equation*}
    \Sigma_j = Proj_{\Sym_3^+}\bigg(\sum_{r=1}^5 \mathsf{FA_{x_j}^r} u_{x_j}^r (u_{x_j}^r)^T \bigg).
\end{equation*}
Often, these covariances are not full rank, so we project them into the set of symmetric positive definite matrices $\Sym_3^+$ by replacing any non-positive eigenvalues with a small positive value. Thus, letting $\theta_j = (x_j, \Sigma_j)$, and normalizing the TDI weights as $b_j = w_j/(\sum_{j=1}^m w_j)$,
we represent the ROI of subject $\ell$ as an empirical probability measure on $\R^3 \times \Sym_3^+$

\begin{figure*}[t]
    \centering
    \includegraphics[width=\textwidth]{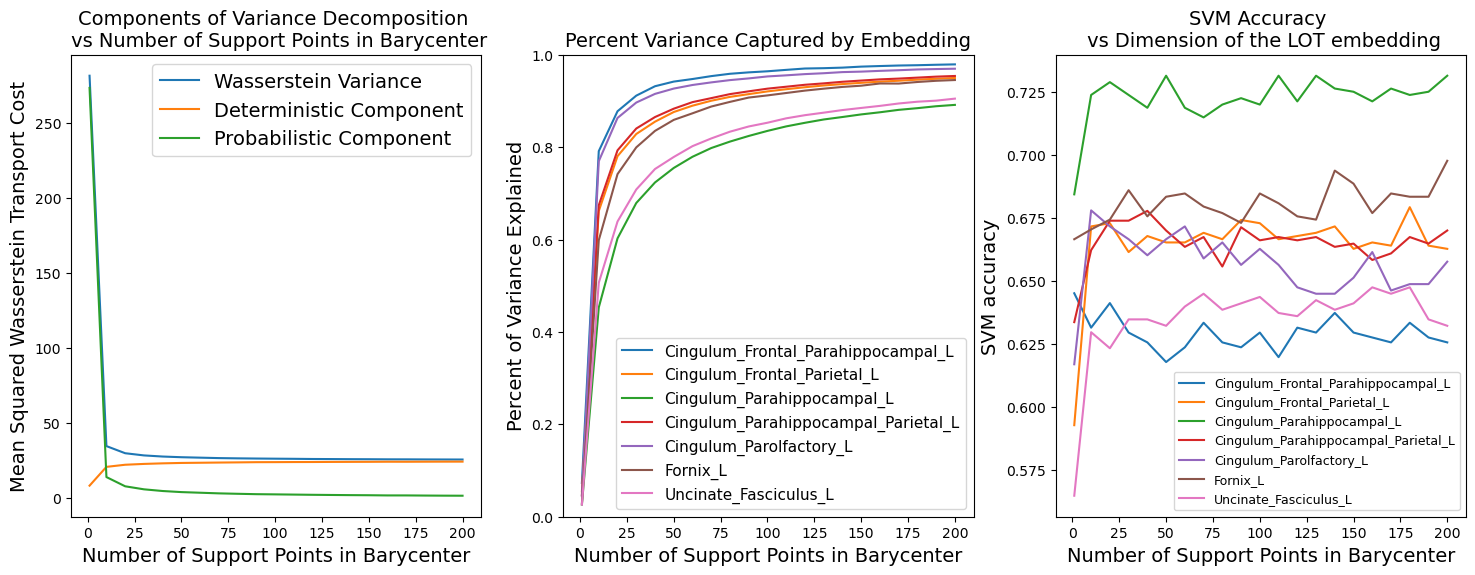}
    \caption{Left: Fr\'echet variance and components of decomposition as a function of number of support points in free support barycenter (for n=1,10,20,...,200) for Fornix L in HCP-YA; Middle: percentage of variance explained by LOT embedding with respect to a free support barycenter with n support points, for ROI (in left hemisphere) in HCP-YA; Right: SVM classification accuracy on LOT embedded data with respect to number of support points in free support barycenter.}
    \label{fig: DTMRI 2}
\end{figure*}

\begin{equation*}
    \mu_\ell = \sum_{j=1}^{m_{\ell}} b_j \theta_j.
\end{equation*}
For the purposes of calculating the decomposition, we conduct our statistical analysis by equipping $\R^3$ and $\Sym_3^+$ with their respective Euclidean metrics. While, at this point, one could choose other Riemannian metrics on $\R^3$ and $\Sym_3^+$ \cite{sarrazin2023linearized}, we choose to use the Euclidean metrics because it enables us to isometrically embed the locations $\theta$ into $\R^9$ (and thus easily calculate the percentage of variance explained by the LOT embedding) using the mapping

\begin{align}
\bigg(   
  & \begin{bmatrix}
    v_{1}  \\
    v_{2}  \\
    v_{3} 
  \end{bmatrix} ,
  \begin{bmatrix}
    v_{4} & v_{5} & v_{6} \\
    v_{5} & v_{7} & v_{8} \\
    v_{6} & v_{8} & v_{9}
  \end{bmatrix}
\bigg) \nonumber\\ 
&\to [v_1,v_2,v_3,v_4,2v_5,2v_6,v_7,2v_8,v_9]^T.
\label{eq: reshape}
\end{align}

\subsubsection{Calculating the LOT Embedding}

Given a set of empirical measures $\mu_\ell, \ell = 1,...,N$, each representing a particular ROI in a subject, we calculate a free support barycenter $\nu$ using the algorithm described in \cite{cuturi2014fast}
(initialized with uniform weights, points in $\R^3$ generated from a Standard Normal distribution, and 3-d covariances generated from a log-uniform distribution \cite{miolane2020geomstats} over $\Sym_3^+$). The free support barycenter calculated for the tract `Fornix L' (with $n=200$ support points) is presented in the left panel of Fig.~\ref{fig: DTMRI 1}.

Next, given $\mu_\ell, \ell = 1,...,N$, and the free support barycenter $\nu = \sum_{i=1}^{n} a_i \theta_i$, we calculate matrix representations of the vector fields of each subject {$V^\ell \in  \R^{n \times d}$} as in Eq.~\eqref{eq: barycentric projection}. The $V^\ell$s can be vectorized as $Z_\ell = Vec(V^\ell) \in \R^{nd}$, at which point, standard statistical methods (i.e. PCA, Regression, SVM) can be applied to the LOT embedded dataset $Z \in \R^{nd \times N}$. The original data and the barycentric projection representation for a particular subject (calculated with respect to the barycenter presented in the left panel of Fig.~\ref{fig: DTMRI 1}) are presented in the middle and right panels of Fig.~\ref{fig: DTMRI 1}, respectively.

\subsubsection{DTMRI Classification Results}

Fig.~\ref{fig: DTMRI 2} presents results of numerical experiments applied to the DTMRI data, similar to those presented in Fig.~\ref{fig:MNIST 2}. The left panel corresponds to the components of the variance decomposition specifically for the `Fornix L.' The middle panel presents elbow plots for all ROI (in the left hemisphere). Notably, we see that for all regions, the large majority of variation is explained by LOT embeddings calculated with respect to free support barycenters with just $25-75$ support points. Given that most regions have on the order of $1000's$ of support points, this corresponds to a significant reduction of dimensionality. The right panel corresponds to the performance of SVM classifiers built on \emph{only the covariance components} (we include results for classifiers built on other combinations of location and covariance components in appendix \ref{app: classification}) of the LOT embeddings with respect to an $n$ support barycenter, for $n=1,10,20,50,100,200$. For most regions, we see a large jump in performance going from 1 support point to $10$ support points, with most models hitting maximum performance (relative to choice of $n$) by about $n=50$ support points. 

Table~\ref{tab: results} presents the results of Leave-One-Out (LOO) cross validation for an SVM classifier applied to only the covariance components of the vector representations of the LOT embedded measures, for different values of $n$ (number of support points in barycenter). We also compare our method to a `baseline' model that is often used in DTMRI analysis, which simply involves comparing the average FA value for each region of each subject \cite{kanaan2012gender, takao2014sex, van2017sex}. Notably, our results suggest that our model outperforms the baseline for all values of $n$, and that SVMs built on LOT embedded data achieve near maximum classification accuracy (relative to other choices of parameters) for free support barycenters with just 10 support points, with little to no improvement for larger values of $n$, both for individual regions and for a soft voting classifier built on class probabilities of the SVMs built on the individual regions, presented in the bottom row of Table~\ref{tab: results}.

\begin{table*}
    \centering
    Classification of Gender in HCP-YA
    \resizebox{\textwidth}{!}{
    \begin{tabular}{| l | c | c c c c c c |}
        \multicolumn{1}{l}{Region} & \multicolumn{1}{c}{N} & \multicolumn{6}{c}{Accuracy (\%)}   \\ 
        \hline
        ~ & ~ & {Average FA} & n=1 &  n=10 &  n=50 & n=100 &  n=200    \\ \hline
        Cingulum Frontal Parahippocampal L & 513 & 63.93 & 64.52 & 62.96 & 62.18 & 62.57 & 62.77\\
        Cingulum Frontal Parahippocampal R & 476 & 63.22 & 65.76 & 65.55 & 65.76 & 64.29 & 65.13\\
        Cingulum Frontal Parietal L & 786 & 53.06 & 59.29 & 66.92 & 66.28 & 66.92 & 65.9\\
        Cingulum Frontal Parietal R & 785 & 59.49 & 63.69 & 68.54 & 68.54 & 68.15 & 68.03\\
        Cingulum Parahippocampal L & 786 & 67.17 & 68.45 & 71.25 & 72.52 & 73.03 & 72.39\\
        Cingulum Parahippocampal R & 786 & 63.36 & 64.76 & 70.1 & 70.74 & 69.59 & 69.85\\
        Cingulum Parahippocampal Parietal L & 770 & 63.38 & 63.38 & 65.97 & 67.4 & 66.1 & 66.49\\
        Cingulum Parahippocampal Parietal R & 779 & 57.38 & 64.57 & 67.27 & 67.39 & 66.5 & 65.85\\
        Cingulum Parolfactory L & 786 & 56.85 & 61.7 & 67.81 & 67.81 & 65.27 & 65.39\\
        Cingulum Parolfactory R & 786 & 57.88 & 58.27 & 60.81 & 62.34 & 62.72 & 62.34\\
        Fornix L & 771 & 65.36 & 66.67 & 66.93 & 68.09 & 69.0 & 68.35\\
        Fornix R & 768 & 63.16 & 64.45 & 67.32 & 67.71 & 68.49 & 68.75\\
        Uncinate Fasciculus L & 786 & 55.99 & 56.49 & 63.23 & 63.61 & 64.38 & 64.38\\
        Uncinate Fasciculus R & 786 & 53.80 &  53.44 & 63.99 & 64.89 & 65.01 & 64.76\\ \hline
        All* Regions Combined & 739 & 64.95 & 69.14 & 75.10 & 75.23 & 74.96 & 75.10 \\ \hline
    \end{tabular}
    }

    \footnotesize{*Cingulum Frontal Parahippocampal L \& Cingulum Frontal Parahippocampal R excluded due to lower sample sizes}

    \caption{Leave One Out Cross Validation Accuracy for SVM built on covariance components of LOT embeddings of DTMRI data with respect to $n$-support barycenters. N denotes sample size - there is missingness due to issues in data preproccessing.}
    \label{tab: results}
\end{table*}

\section{Conclusion}\label{Conclusion}

This paper presents a decomposition of the Fused 2-Gromov-Wasserstein Fr\'echet variance of measures supported on Euclidean spaces using Linear Optimal Transport. The decomposition gives data scientists a tool that aids in parameter selection when working with $FGW$ distances and helps them better interpret the quality of LOT embeddings. It also suggests connections to other methods for calculating Free Support barycenters \cite{luise2019sinkhorn}. The decomposition also suggests a generalization of the F-statistic used in ANOVA, which allows one to test a data set of empirical probability measures for equality of $n$ support barycenter. Furthermore, our methods have provided promising results for the statistical modeling of DTMRI images, using Linear Optimal Transport to naturally fold dimensionality reduction and non-linear image registration into standard machine learning pipelines. This application suggests many directions for future research, such as increasing the complexity of the data representations by using other Riemannian metrics on SPD matrices \cite{sarrazin2023linearized} or `orientation distribution functions' rather than covariances \cite{dhollander2014track}, scaling OT with convolutions and/or Sinkhorn \cite{cuturi2013sinkhorn,feydy2019fast} to build models on larger regions or the whole brain, and/or implementing partial/unbalanced LOT methods \cite{bai2023linear}.  

\subsection{Acknowledgments}

This research was supported in part by NSF grants DMS-2107808, DMS-2324962 to Needham and DMS-1953087, DMS-241378 to Srivastava, and NIH grant R01 MH120299 to Srivastava.
DTMRI data were provided [in part] by the Human Connectome Project, WU-Minn Consortium (Principal Investigators: David Van Essen and Kamil Ugurbil; 1U54MH091657) funded by the 16 NIH Institutes and Centers that support the NIH Blueprint for Neuroscience Research; and by the McDonnell Center for Systems Neuroscience at Washington University.

\bibliographystyle{plain}
\bibliography{refs}

\newpage

\appendix

\section{Proofs}

\subsection{Proof of Proposition \ref{prop: w decomp}}\label{app: w decomp}
    
\begin{proof}
First note that
\begin{align}\label{test}
    &  W_2^2(\nu, \mu) = \sum_{ij} \gamma^*_{ij}\|x_i-y_j\|^2 \nonumber \\
    & =  \sum_{ij} \gamma^*_{ij} \|(x_i-T^*(x_i)) + (T^*(x_i) -y_j)\|^2 \nonumber \\
    & =  \sum_{i} a_i \|x_i - T^*(x_i)\|^2 + \sum_{ij} \gamma^*_{ij} \|T^*(x_i) - y_j\|^2 \\
    &- 2  \sum_{ij} \gamma^*_{ij} \langle x_i-T^*(x_i), T^*(x_i) - y_j \rangle \nonumber \\
    & =  \sum_{ij} a_{i} \|x_i - T^*(x_i)\|^2 +  \sum_{ij} \gamma^*_{ij} \|T^*(x_i) - y_j\|^2 
\end{align}
where the last equality follows because
\begin{align*}
    & \sum_{ij} \gamma^*_{ij} \langle x_i-T^*(x_i), T^*(x_i) - y_j \rangle\\
    & = \sum_i a_i \bigg( \sum_j \frac{\gamma^*_{ij}}{a_i} \langle x_i-T^*(x_i), T^*(x_i) - y_j \rangle  \bigg)  \\
    & = \sum_i a_i \bigg( \langle x_i-T^*(x_i), T^*(x_i) - \sum_j \frac{\gamma^*_{ij}}{a_i} y_j \rangle  \bigg)  = 0.
\end{align*}
We still have to show that 
\[
W_2^2(\nu,T^*_\#\nu) = \sum_i a_i \|x_i - T^*(x_i)\|^2.
\]
First observe that, using the transport cost functions defined in Definition~\ref{def: transport cost}, Eq.~\ref{test} implies that 
\begin{align*}
    & C_W(\gamma^*) =  \sum_{ij} a_{i} \|x_i - T^*(x_i)\|^2 +  \sum_{ij} \gamma^*_{ij} \|T^*(x_i) - y_j\|^2 \\
    & = C_W(T^*) + \sum_{ij} \gamma^*_{ij}\|T^*(x_i)\|^2 - 2 \langle T^*(x_i), y_j \rangle +  \|y_j\|^2 \\
    & = C_W(T^*) +  \sum_{i} a_i\|T^*(x_i)\|^2 - 2 \sum_{ij} \gamma^*_{ij}\langle T^*(x_i), y_j \rangle\\
    &\hspace{2in}+ \sum_{j} b_j \|y_j\|^2 \\
    & = C_W(T^*) +   \sum_{i} a_i\|T^*(x_i)\|^2\\
    &\qquad \qquad - 2 \sum_{i} a_i \langle T^*(x_i), \sum_{j} \frac{\gamma^*_{ij}}{a_i} y_j \rangle + \sum_{j} b_j \|y_j\|^2 \\
    & = C_W(T^*) +  \bigg(\sum_j b_j \|y_j\|^2 - \sum_{i} a_i \|T^*(x_i)\|^2\bigg).
\end{align*}
Let $\hat{\gamma}$ be an arbitrary element of $\Pi(\nu, T_\#\nu)$, and (following the approach used in the proof of Lemma 2.1 in \cite{bai2023linear}) define a coupling $\gamma \in \Pi(\nu, \mu)$ such that $\gamma_{ij} = \sum_{p=1}^n \frac{\hat{\gamma}_{ip} \gamma^*_{pj}}{a_p}$. Now observe 
\begin{align*}
    & C_W(\gamma) = \bigg(\sum_{i} a_i \|x_i\|^2 +  \sum_{j} b_j \|y_j\|^2\bigg) \\
    &\hspace{1.5in} - 2\sum_{i=1}^n \sum_{j=1}^m \gamma_{ij}  \langle x_i, y_j \rangle\\
    & = \bigg(\sum_{i} a_i \|x_i\|^2 +  \sum_{j} b_j \|y_j\|^2\bigg) \\
    &\hspace{1.5in} - 2\sum_{i=1}^n \sum_{j=1}^m \sum_{p=1}^n \frac{\hat{\gamma}_{ip} \gamma^*_{pj}}{a_p}  \langle x_i, y_j \rangle\\
    & = \bigg(\sum_{i} a_i \|x_i\|^2 +  \sum_{j} b_j \|y_j\|^2\bigg) \\
    &\hspace{1.5in} - 2\sum_{i=1}^n \sum_{p=1}^n  \hat{\gamma}_{ip} \langle x_i, T^*(x_p) \rangle\\
    & = C_W(\hat{\gamma}) + \bigg(\sum_j b_j \|y_j\|^2 - \sum_{p} a_p \|T^*(x_p)\|^2\bigg)
\end{align*}
Thus,
\begin{align*}
    &C_W(\gamma^*) \leq C_W(\gamma) \\
    &\iff C_W(T^*) \leq C_W(\hat{\gamma}) \ \forall \hat{\gamma} \in \Pi(\nu, T^*_\#\nu)\\
    &\implies W_2^2(\nu,T^*_\#\nu) = \sum_i a_i \|x_i - T^*(x_i)\|^2
\end{align*}
which, combined with Eq.~\ref{test}, gives the desired result.
\end{proof}

\subsection{Proof of Theorem \ref{thm: gw decomp}}\label{app: gw decomp}

    \begin{proof}    
        First note that,
        \begin{align}\label{test2}
            & \sum_{ijkl} \gamma^*_{ij} \gamma^*_{kl} |A_{ik} - B_{jl}|^2 \nonumber \\
            & = \sum_{ijkl} \gamma^*_{ij} \gamma^*_{kl} |A_{ik} - C^*_{ik} + C^*_{ik} - B_{jl}|^2 \nonumber\\
            & = \sum_{ijkl} \gamma^*_{ij} \gamma^*_{kl} \big( |A_{ik} - C^*_{ik}|^2 + |C^*_{ik} - B_{jl}|^2 \big) \nonumber\\
            & + \sum_{ijkl} \gamma^*_{ij} \gamma^*_{kl}  2(A_{ik} - C^*_{ik})(C^*_{ik} - B_{jl}) \nonumber\\
            & = \sum_{ik} a_i a_k|A_{ik} - C^*_{ik}|^2 + \sum_{ijkl} \gamma^*_{ij} \gamma^*_{kl} |C^*_{ik} - B_{jl}|^2
        \end{align}       
        where the last equality follows because
            
        \begin{align*}
            & \sum_{ijkl} \gamma^*_{ij} \gamma^*_{kl} (A_{ik} - C^*_{ik})(C^*_{ik} - B_{jl}) \\
            & = \sum_{ik} a_i a_k (A_{ik} - C^*_{ik}) \bigg( \sum_{jl} \frac{\gamma^*_{ij}}{a_i} \frac{\gamma^*_{kl}}{a_k}(C^*_{ik} - B_{jl}) \bigg) \\
            & = \sum_{ik} a_i a_k (A_{ik} - C^*_{ik}) \bigg(C^*_{ik} - \sum_{jl} \frac{\gamma^*_{ij}}{a_i} \frac{\gamma^*_{kl}}{a_k} B_{jl} \bigg) = 0.
        \end{align*}
We still have to show that $GW_2^2(\mathcal{X},\mathcal{T}) = \sum_{ik} a_i a_k|A_{ik} - C^*_{ik}|^2$. First observe that, using the transport cost functions defined in Definition~\ref{def: transport cost}, Eq.~\ref{test2} implies that 

     \begin{align*}
        & C_{GW}(\gamma^*) = \sum_{ik} a_i a_k|A_{ik} - C^*_{ik}|^2 + \sum_{ijkl} \gamma^*_{ij} \gamma^*_{kl} |C^*_{ik} - B_{jl}|^2 \\
        & = C_{GW}(T^*) + \sum_{ijkl} \gamma^*_{ij} \gamma^*_{kl} {C^*_{ik}}^2  - 2 C^*_{ik} B_{jl} +  B_{jl}^2 \\
        & = C_{GW}(T^*) + \sum_{ik} a_i a_k {C^*_{ik}}^2\\
        &\qquad \qquad - 2 \sum_{ik} a_i a_k C^*_{ik} ( \sum_{jl}\frac{\gamma^*_{ij}}{a_i} \frac{\gamma^*_{kl}}{a_k} B_{jl} ) + \sum_{jl} b_j b_l B_{jl}^2 \\
        & = C_{GW}(T^*) + \bigg( \sum_{jl} b_j b_l B_{jl}^2 - \sum_{ik} a_{i} a_k {C^*_{ik}}^2 \bigg)
    \end{align*}
Let $\hat{\gamma}$ be an arbitrary element of $U(a,a)$, and (following the approach used in the proof of Lemma 2.1 in \cite{bai2023linear}) define $\gamma_{ij} = \sum_{p} \frac{\hat{\gamma}_{ip} \gamma^*_{pj}}{a_p}$, noting that $\gamma \in U(a,b)$ since

\begin{align*}
    & \sum_j \gamma_{ij} = \sum_p \frac{\hat{\gamma}_{ip}}{a_p} \sum_j \gamma^*_{pj} = \sum_p \hat{\gamma}_{ip} = a_i\\
    & \sum_i \gamma_{ij} = \sum_p \frac{\gamma^*_{pj}}{a_p} \sum_i \hat{\gamma}_{ip} = \sum_p \gamma^*_{pj} = b_j
\end{align*}
Now observe 

\begin{align*}
    C_{GW}(\gamma) & = \bigg(\sum_{ik} a_i a_k A_{ik}^2 +  \sum_{jl} b_j b_l B_{jl}^2\bigg)\\
    &\qquad \qquad - 2\sum_{ijkl} \gamma_{ij} \gamma_{kl} A_{ik} B_{jl}\\
    & = \bigg(\sum_{ik} a_i a_k A_{ik}^2 +  \sum_{jl} b_j b_l B_{jl}^2\bigg)\\
    &\qquad \qquad - 2\sum_{ijkl} \sum_{pq} \frac{\hat{\gamma}_{ip} \gamma^*_{pj}}{a_p} \frac{\hat{\gamma}_{kq} \gamma^*_{ql}}{a_q}  A_{ik}B_{jl}\\
    & = \bigg(\sum_{ik} a_i a_k A_{ik}^2 +  \sum_{jl} b_j b_l B_{jl}^2\bigg)\\
    &\qquad \qquad - 2\sum_{ipkq}  \hat{\gamma}_{ip} \hat{\gamma}_{kq} A_{ik}C^*_{pq}\\
    & = C_{GW}(\hat{\gamma}) + \bigg(\sum_{jl} b_j b_l B_{jl}^2 - \sum_{pq} a_p a_q (C^*_{pq})^2\bigg)
\end{align*}
Thus,
\begin{align*}
    & C_{GW}(\gamma^*) \leq C_{GW}(\gamma) \\
    &\iff C_{GW}(T^*) \leq C_{GW}(\hat{\gamma}) \ \forall \hat{\gamma} \in U(a,a)\\
    & \implies GW_2^2(\nu,\tilde{\mu}) = \sum_{ik} a_i a_k|A_{ik} - C^*_{ik}|^2
\end{align*}
which, combined with Eq.~\ref{test2}, gives the desired result.
\end{proof}

\subsection{Other Classification Results}\label{app: classification}

In practice, it may be of scientific interest to test hypotheses that consider only diffusion information, only location information, or some combination of both. For this, one can use a scalar $\lambda \in [0,1]$ to weigh the location versus covariance components, that is, one can apply the embedding in \eqref{eq: reshape} instead to coordinates where the vectors in $\R^3$ are scaled by $\lambda$, and the matrices in $\Sym_3^+$ are scaled by $(1-\lambda)$. For this experiment, we use $n=200$, and $\lambda=0,\lambda^*,1$, where, letting $Z_\lambda$ denote the LOT embedded dataset with parameter $\lambda$ used in the vectorization,

\begin{equation*}
    \lambda^* \coloneqq \frac{tr\big((Z_0)^T(Z_0)\big)}{tr\big((Z_1)^T(Z_1)\big)+tr\big((Z_0)^T(Z_0)\big)}.
\end{equation*}
is intended to balance the variances of the location and covariance components. Table~\ref{second experiment} contains the results for these experiments. 

\begin{table}[t]
    \centering
    Classification of Gender in HCP-YA
    \resizebox{.5\textwidth}{!}{
    \begin{tabular}{| l | c | c c c | c |}
        \multicolumn{1}{l}{Region} & \multicolumn{1}{c}{N} & \multicolumn{3}{c}{Accuracy (\%)}   \\ 
        \hline
        ~ & N &  $\lambda = 0$ &  $\lambda = \lambda^*$ &  $\lambda = 1$   \\ \hline
        Cingulum Frontal Parahippocampal L & 513 & 65.69 & 65.88 & 59.84   \\ %\hline
        Cingulum Frontal Parahippocampal R & 476 & 65.33& 62.81 & 56.51  \\ %\hline
        Cingulum Frontal Parietal L & 786 & 66.66& 65.39 & 57.88 \\ %\hline
        Cingulum Frontal Parietal R & 785 & 65.98& 66.11 & 59.61 \\ %\hline
        Cingulum Parahippocampal L & 786 & 71.37& 73.28 & 61.06 \\ %\hline
        Cingulum Parahippocampal R & 786 & 70.48 & 71.75 & 61.06  \\ %\hline
        Cingulum Parahippocampal Parietal L & 770 & 66.10 & 66.88 & 59.87  \\ %\hline
        Cingulum Parahippocampal Parietal R & 779 & 66.23& 64.31 & 59.30  \\ %\hline
        Cingulum Parolfactory L & 786 & 65.52 & 65.77 & 54.32  \\ %\hline
        Cingulum Parolfactory R & 786 & 63.23 & 65.13 & 56.23 \\ %\hline
        Fornix L & 771  & 69.90 & 70.55 & 55.25  \\ %\hline
        Fornix R & 768  & 69.66 & 68.09 & 55.46  \\ %\hline
        Uncinate Fasciculus L & 786 & 65.26 & 62.84 & 60.68  \\ %\hline
        Uncinate Fasciculus R & 786 & 61.57 & 60.55 & 61.70 \\ \hline
        All* Regions Combined & 739 & 74.96 & 74.01 & 69.01\\ \hline
    \end{tabular}
    }

    \footnotesize{*Cingulum Frontal Parahippocampal L \& Cingulum Frontal Parahippocampal R excluded due to lower sample sizes}

    \caption{Leave One Out Cross Validation Accuracy for SVM built LOT embedding with respect to a 200 support point barycenter, for each region, for different values of $\lambda$. N denotes sample size - There is missingness due to the fact that DSI studio does not return results for some regions of some subjects.}

    \label{second experiment}
    
\end{table}

\section{Algorithms}\label{sec: algorithms}

\subsection{Gaussian Kernel Reconstruction for MNIST images}\label{app: gaussian kernel}

Given an empirical probability measure $\nu = \sum_{i=1}^n a_i \delta_{x_i}$ supported on $[0,27]^2$, we define an empirical probability measure with locations $ (i,j) \in \{0,...,27\}^2$ and weights $w(i,j) = A(i,j)/\sum_{ij}A(i,j)$ where $A(i,j) = \sum_{i=1}^n a_i N(x_i|(i,j), I(2))$ where $I(2)$ is the identity matrix in $R^{2 \times 2}$.

\end{document}